# Solving integral equations in free-space with inverse-designed ultrathin optical metagratings

Andrea Cordaro[1,2*], Brian Edwards[3], Vahid Nikkhah[3], Andrea Alù[4,5], Nader Engheta[3], Albert Polman[2]

[1] Institute of Physics, University of Amsterdam, Science Park 904, 1098 XH Amsterdam, The Netherlands
[2] Center for Nanophotonics, NWO-Institute AMOLF, Science Park 104, 1098 XG Amsterdam, The Netherlands
[3] Department of Electrical and Systems Engineering, University of Pennsylvania, Philadelphia, PA 19104, USA.
[4] Photonics Initiative, Advanced Science Research Center, City University of New York, New York, NY 10031, USA
[5] Physics Program, Graduate Center, City University of New York, New York, NY 10016, USA
[*] e-mail: a.cordaro@amolf.nl

**As standard microelectronic technology approaches fundamental limitations in speed and power consumption, novel computing strategies are strongly needed. Analog optical computing enables processing large amounts of data at a negligible energy cost and high speeds. Based on these principles, ultrathin optical metasurfaces have been recently explored to process large images in real-time, in particular for edge detection. By incorporating feedback, it has also been recently shown that metamaterials can be tailored to solve complex mathematical problems in the analog domain, although these efforts have so far been limited to guided-wave systems and bulky setups. Here, we present an ultrathin Si metasurface-based platform for analog computing that is able to solve Fredholm integral equations of the second kind using free-space visible radiation. A Si-based metagrating was inverse-designed to implement the scattering matrix synthesizing a prescribed Kernel corresponding to the mathematical problem of interest. Next, a semi-transparent mirror was incorporated into the sample to provide adequate feedback and thus perform the required Neumann series, solving the corresponding equation in the analog domain at the speed of light. Visible wavelength operation enables a highly compact, ultrathin device that can be interrogated from free-space, implying high processing speeds and the possibility of on-chip integration.**

The world's ever-growing needs for efficient computing have been driving researchers from diverse research fields to explore alternatives to the current digital computing paradigm. The processing speed and energy efficiency of standard electronics have become limiting factors for novel disruptive applications entering our everyday life, such as artificial intelligence, machine learning, computer vision, and many more. In this context, analog computing has resurfaced and regained significant attention as a complementary route to traditional architectures[1–4]. Specifically, the tremendous recent advances in the field of metamaterials and metasurfaces have been unlocking new opportunities for all-optical computing strategies, given the possibility of shaping optical fields in extreme ways over subwavelength thicknesses. The absence of bulky optical elements, in turn, enables on-chip integration paving the way for hybrid optical and electronic data processing.

The idea of using light to outsource specific computing tasks comes with several advantages. First, there is a clear enhancement in processing speeds as the computation is performed at the speed of light traveling through metamaterials with typical sizes smaller than or comparable with the wavelength of operation. Also, processing signals in the optical domain enables massive parallelization and may potentially avoid unnecessary analog-to-digital conversion. As an example, recent works have shown how several image processing tasks can be performed before the image is discretized into pixels[5–19], relying on the possibility of engineering the angular response of metasurfaces and hence impart instantaneously a mathematical operation to the spatial content of an input signal[11,12,20].

Finally, analog computing meta-devices can be passive, implying an extremely low energy usage. Recently, broader applications of this approach have been appearing in different fields, ranging from silicon photonics[21–23] to organic neuromorphic electronics,[24,25] i.e., architectures that mimic the biological brain's function, and even acoustics.[26–28]

A key question is whether it is possible to go beyond simple image processing tasks and focus on a more complex mathematical problem, such as solving an *integro-differential equation*. The concept of a wave-based integral equation solver has been recently demonstrated in the microwave regime for symmetric and non-symmetric kernels and in a multi-frequency parallel fashion[29,30], but relying on guided waves in bulky metamaterial setups. An important next challenge is to demonstrate if such a complex mathematical operation can be carried out in the optical spectral range, ideally within an ultrathin form factor that can be interrogated through free-space radiation and easily combined with similar devices to represent operator composition. This will enable the fabrication of far more compact on-chip devices operated at wavelengths that are widely used for communication technology. This dramatic size reduction further implies a drastic increase in processing speeds as light has to travel much shorter distances.

Here, we demonstrate a Si metasurface-based optical platform that combines a tailored scattering matrix design and a feedback system to enable the solution of Fredholm integral equations of the second kind from the far-field

$$g(u) = I_{in}(u) + \int_a^b K(u,v)\, g(v) dv, \quad (1)$$

where $g(u)$ is the unknown solution of Eq. (1), $K(u,v)$ is the kernel of the integral operator, and $I_{in}(u)$ is an arbitrary input function. Mathematically, this form of equation may be analytically solvable if it is in separable form or for some special kernels, and an inversion formula may exist (e.g., a Fourier transform). However, when certain convergence conditions for the kernels are satisfied a general technique to solve Eq. (1) is to exploit the Neumann successive approximation method: we assume an initial guess $g_0(u) = I_{in}(u)$ and successive approximations can be obtained by evaluating $g_{i+1}(u) = I_{in}(u) + \int_a^b K(u,v)\, g_i(v) dv$, whereupon eventually $g_n(u)$ converges to the solution $g(u)$ as $n \to \infty$.[31] Here, we show how to physically implement this iterative procedure in an analog fashion employing a Si metasurface coupled to a feedback system.

First, Eq. (1) is discretized by sampling its independent variables, u and v, over *N* points in the interval $[a,b]$ to form two vectors with such variables $\underline{u}$ and $\underline{v}$. The application of the integral operator $\int_a^b K(u,v)\,[\ ]\, dv$ on the function $g(u)$ is then analogous to the multiplication (or application) of a matrix operator $\mathbf{K} = K(\underline{u},\underline{v})\frac{(a-b)}{N}$ on a vector $\underline{g} = g(\underline{u})$. Thus, Eq. (1) may be numerically approximated by the $N \times N$ matrix equation

$$\underline{g} = \underline{I_{in}} + \mathbf{K}\,\underline{g} \quad (2)$$

Second, the solution $\underline{g}$ is represented as a Neumann series $\underline{g} = \sum_n (\mathbf{K})^n\, \underline{I_{in}} = (\mathbf{I_N} - \mathbf{K})^{-1} \underline{I_{in}}$, where $\mathbf{I_N}$ is the $N \times N$ identity matrix. The convergence of the Neumann series demonstrates that the inverse operator $(\mathbf{I_N} - \mathbf{K})^{-1}$ exists.

Next, it is possible to think of the *N* mathematical sampling points as *N* discrete physical modes, and thus $\underline{g}$ is a vector representing the complex amplitude of these modes on a given plane with a chosen direction. The integral operator can then be represented by a scattering matrix that performs matrix multiplication between these sets of modes.

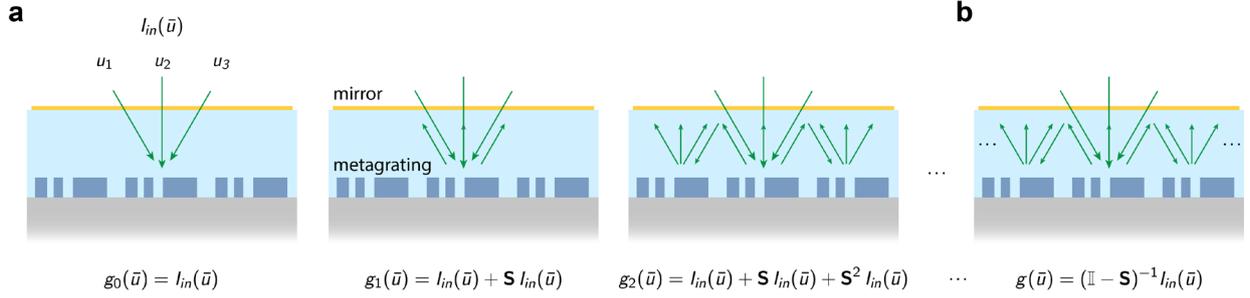

Fig. 1 | **Si metagrating-based integral equation solver.** An input vector $\underline{I_{in}}$ is fed to the system in the form of N plane waves with different complex amplitudes incident along N diffraction channels. The signal interacts repeatedly with a metagrating bouncing back from a partially reflecting mirror, each time multiplied by the metagrating scattering matrix and therefore building up the terms of a Neumann series of subsequent matrix multiplications required for solving the integral equation. For the sake of simplicity, the formulas underneath the panels do not take into account the semi-transparent mirror scattering matrix at this stage.

If we consider a periodic metagrating, the input/output modes can be mapped into the *N* discrete diffraction channels determined by the periodicity and the wavelength, while the discretized integral operator **K** can be mapped onto the metasurface scattering matrix **S** that governs the coupling between these channels. Following the schematics in Fig. 1, the discretized input $\underline{I_{in}}$ is a vector of length *N* containing the complex amplitudes of the plane waves addressing the system via its available diffraction channels, acting as seed guess $\underline{g_0} = \underline{I_{in}}$. The vector is multiplied by the metasurface scattering matrix upon its first reflection, resulting in a more refined guess $\underline{g_1}$ to the solution of the integral equation associated with **K**. The signal is then reflected by a semitransparent mirror and fed back to the grating for the next iteration. Intuitively, the system performs an analog Neumann series at the speed of light by iteratively applying the **S** matrix on the seed vector through multiple reflections, in the same way the mathematical integral operator is applied repeatedly on the initial guess function.

The entire computing metastructure is therefore composed of two elements: (1) a metagrating with a period that determines the number of input/output modes (grating orders), and unit cell with tailored geometry defining the scattering matrix of interest, (2) a semi-transparent mirror enabling feedback and in-coupling combined with a spacer layer.

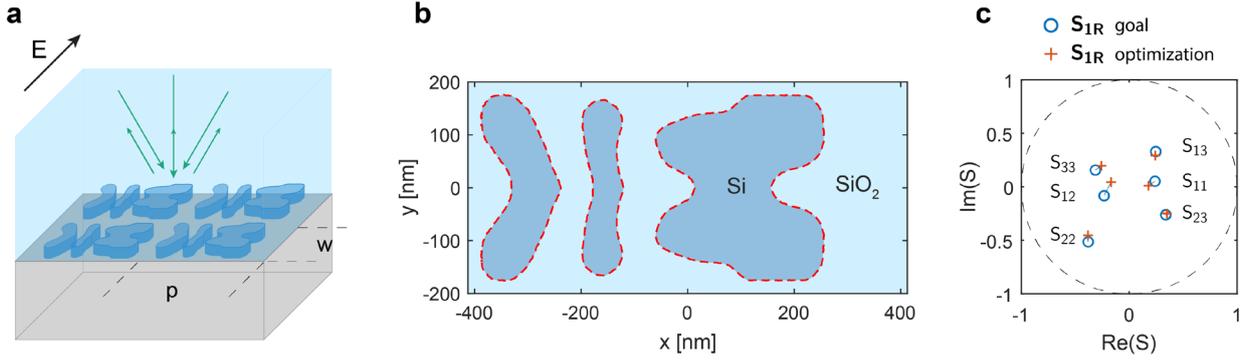

Fig. 2 | **Kernel design. a** Schematic illustration of a 2D metagrating (periodicities *p* and *w*) with a suitably engineered unit cell geometry. The black arrow indicates the polarization (TE) of the incoming E field. **b** Top view of the optimum metagrating unit cell made of Si (blue) and SiO$_2$ (light blue). **c** Simulation results for the $\mathbf{S_{1R}}$ matrix elements of the inverse-designed metagrating (orange crosses) and the corresponding desired matrix elements (blue circles).

The key requirement is the careful design of the metagrating unit cell to synthesize the prescribed S-matrix

$$\mathbf{S_1} = \begin{pmatrix} \mathbf{S_{1R}} & \mathbf{S_{1T}}^\top \\ \mathbf{S_{1T}} & \mathbf{S_{1R'}} \end{pmatrix} \quad (3)$$

where $\mathbf{S_{1R}}$ and $\mathbf{S_{1T}}$ are the reflection and transmission sub-blocks. Since the Neumann series is performed in reflection, only the reflection sub-block $\mathbf{S_{1R}}$ needs to be designed: it contains the complex reflection coefficients connecting the diffraction channels in the spacer layer above the metagrating. To prove the generality of our approach, we began our problem by choosing a random passive and reciprocal (i.e., symmetric) matrix with *N*=3.

$$\mathbf{S_{1R}} = \begin{pmatrix} 0.239 + 0.052\,i & -0.233 - 0.083\,i & 0.246 + 0.329\,i \\ -0.233 - 0.083\,i & -0.381 - 0.514\,i & 0.339 - 0.262\,i \\ 0.246 + 0.329\,i & 0.339 - 0.262\,i & -0.314 + 0.156\,i \end{pmatrix} \quad (4)$$

Next, we set the periodicities *p* and *w* of the grating (see Fig. 2a) to have three diffraction orders in reflection at the target wavelength $\lambda_0$ = 706 nm. We choose this wavelength because (1) Si is rather transparent, (2) light sources in this spectral range are readily accessible, and (3) the corresponding submicron unit cell footprint enables compact circuit design and integration. Specifically, we choose *p* = 825 nm while the orthogonal periodicity *w* = 400 nm is set to be subwavelength. This enhances the degrees of freedom for the unit cell design without opening additional diffraction channels. We optimize the metagrating unit cell geometry using the adjoint method[32–36] setting the height *h* of the etched silicon nanostructure to 150 nm. The resulting optimized unit cell (Fig. 2b) consists of a Si nanostructure on a sapphire substrate embedded in a transparent SiO$_2$ spacer layer[37–39]. The figure-of-merit to be minimized during the optimization is the sum of the squared "distances" on the complex plane between the complex-valued matrix elements of the S matrix of a designed geometry (as in Fig. 2b) and the prescribed ones in Eq. (4): $\mathrm{FOM} = \sum_{i,j} \left| \mathbf{S_{1R}}_{ij} - \mathbf{S_{1R\_opt}}_{ij} \right|^2$ (see Supplementary materials).

As shown in Fig. 2c, the optimized metagrating approximates very well the desired S matrix, achieving a figure of merit as low as 0.058. This demonstrates that it is possible to inverse-design metagratings with a prescribed S matrix, showing the feasibility of this optical computing concept for the solution of integral equations with a wide range of kernels.

What is discussed so far concerns only the design of the metagrating scattering matrix mapping the discretized integral Kernel operator **K** in Eq.(2). Next, to find the solution of the integral equation it is crucial to have a feedback system that repeatedly returns the signal reflected from the metagrating back to it so that the Neumann series is constructed. To this end, the SiO$_2$ spacer is covered with a 15 nm thick Au layer to form a semitransparent mirror (Fig. 3a) [40]. The distance between the metagrating and mirror is 487 nm $\approx \lambda_0/n_{SiO2}$ to avoid near-field coupling, which may introduce additional modes into the system. The **S** matrix characterizing the mirror is

$$\mathbf{M} = \begin{pmatrix} \mathbf{M}_R & \mathbf{M}_T^\top \\ \mathbf{M}_T & \mathbf{M}_{R\prime} \end{pmatrix} \quad (5)$$

where $\mathbf{M}_{R\prime}$ is the sub-block representing reflection from the SiO$_2$ side and $\mathbf{M}_T$ is its transmission counterpart. Including the mirror, the scattering matrix of the entire meta-structure (grating, SiO$_2$ spacer, and mirror) becomes

$$\mathbf{S}_2 = \begin{pmatrix} \mathbf{S}_{2R} & \mathbf{S}_{2T}^\top \\ \mathbf{S}_{2T} & \mathbf{S}_{2R\prime} \end{pmatrix} \quad (6)$$

The Neumann series, and thus the solution of Eq. (2), is embedded in $\mathbf{S}_2$. The transmission of the entire stack, as measured in our experiment, is composed of a sum of terms each corresponding to an increasing number of interactions with the metagrating (see Figs. 1, 3a)[41]:

$$\mathbf{S}_{2T} = \mathbf{S}_{1T}\mathbf{M}_T + \mathbf{S}_{1T}\mathbf{M}_{R\prime}\mathbf{S}_{1R}\mathbf{M}_T + \mathbf{S}_{1T}(\mathbf{M}_{R\prime}\mathbf{S}_{1R})^2\mathbf{M}_T + \ldots = \mathbf{S}_{1T}(\mathbf{I}_3 - \mathbf{M}_{R\prime}\mathbf{S}_{1R})^{-1}\mathbf{M}_T. \quad (6)$$

The transmission sub-block $\mathbf{S}_{2T}$ is composed of the inverse operator $(\mathbf{I}_3 - \mathbf{M}_{R\prime}\mathbf{S}_{1R})^{-1}$ solving Eq. (2) multiplied by the mirror transmission $\mathbf{M}_T$ and by the metasurface transmission $\mathbf{S}_{1T}$. In other words, light is coupled into the system passing through the mirror first, and then the solution is outcoupled via the metasurface. Hence, to extract the solution computed by the metastructure, *i.e.* the linear combination of complex amplitudes of the diffracted modes inside the spacer layer that converges after multiple passes, $\mathbf{M}_T$ and $\mathbf{S}_{1T}$ must be de-embedded from $\mathbf{S}_{2T}$. Figure 3b compares the solution $\mathbf{S}_{1T}^{-1}\mathbf{S}_{2T}\mathbf{M}_T^{-1}$ provided by the simulated metastructure transmission to the ideal solution of Eq. (2) with $K = \mathbf{M}_{R\prime}\mathbf{S}_{1R}$ and $\underline{I}_{in}$ equal to the vectors belonging to the canonical basis generating the space of all possible input vectors (i.e., $(1,0,0)^\top, (0,1,0)^\top, (0,0,1)^\top$). Any input vector can be expressed as a linear combination of these, and given the linearity of the metasurface, agreement in the response for these basic excitations ensures that the structure can solve the integral equation problem for arbitrary inputs.

The metasurface-based analog solution and the ideal solution show good agreement for all the inputs, both in terms of the real and imaginary parts. Minor discrepancies are ascribed to the small difference between the desired S matrix and the optimized one (see Fig. 2c) and this result demonstrates that it is possible to design the desired kernel **K** and invert $(\mathbf{I}_N - \mathbf{K})$ in a fully analog fashion.

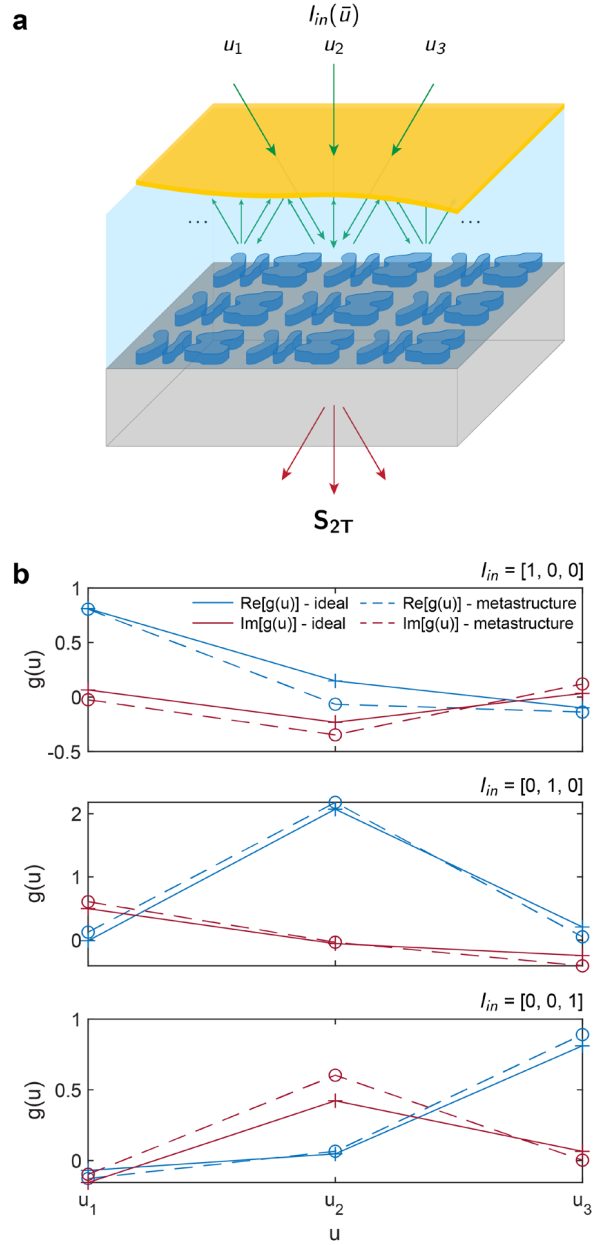

Fig. 3 | **Analog matrix inversion. a** The solution of Eq. (2) is built up inside the spacer layer in the form of a collection of complex wave amplitudes, one for each discrete diffraction channel. The solution is outcoupled and read out in transmission. **b** Analog solution (real and imaginary parts) of the integral equation (dashed line) obtained from the simulation results for the metastructure transmission, compared with the ideal theoretical solution $\underline{g} = (\mathbf{I}_N - \mathbf{K})^{-1}\underline{I_{in}}$ (solid lines), for the three orthogonal input vectors $(1,0,0)^T, (0,1,0)^T, (0,0,1)^T$. The wavelength of operation in this simulation is $\lambda_0$ = 706 nm.

Next, we present the experimental implementation of an all-optical integral equation solving metasurface using the optimized geometry described above. The analog solution of Eq. (2) is built up inside the spacer layer in the form of a collection of complex wave amplitudes. Despite the fact that the complex amplitudes readily exist just below the surface of the kernel and can be utilized by another similar device as the one presented herein, these values are hard to retrieve in the far-field where a meaningful phase reference at each angle is difficult to define. Hence, similar to spectral reflectometry (SR), we obtain a more robust measurement by comparing the spectroscopic power measurements over a broad wavelength range to simulations of the optimized structure. Within SR a limited number of chosen parameters such a material layer thickness, Lorentz oscillator frequency, damping coefficients, etc., are fitted to a vastly overdetermined system to obtain Kramers-Kronig safe models from which any value can be derived, including complex amplitudes with a material stack at a specific frequency. Here we do similar, and allow ourselves only one fitted parameter, a dilation operation on the structure, possibly representing fabrication complications, to generate internally consistent spectral traces for all of the possible inputs. We show that minor perturbations of this single parameter are enough to generate an excellent fit and use this data to retrieve an estimate for the experimental solution provided by the metastructure.

First, the optimized metagrating geometry was patterned over a 150-nm-thick Si(100) film on a sapphire ($Al_2O_3$) substrate by means of electron beam lithography (EBL) and reactive ion etching (RIE). Next, the metasurface was embedded in $SiO_2$ by spin coating and annealing a silica glass sol-gel layer that planarizes the structure, followed by a final $SiO_2$ sputtering that allows fine control of the total spacer thickness. Finally, a 15-nm-thick Au layer was evaporated on the structure using an organic adhesion monolayer (details concerning the fabrication can be found in Supplementary materials)[42,43].

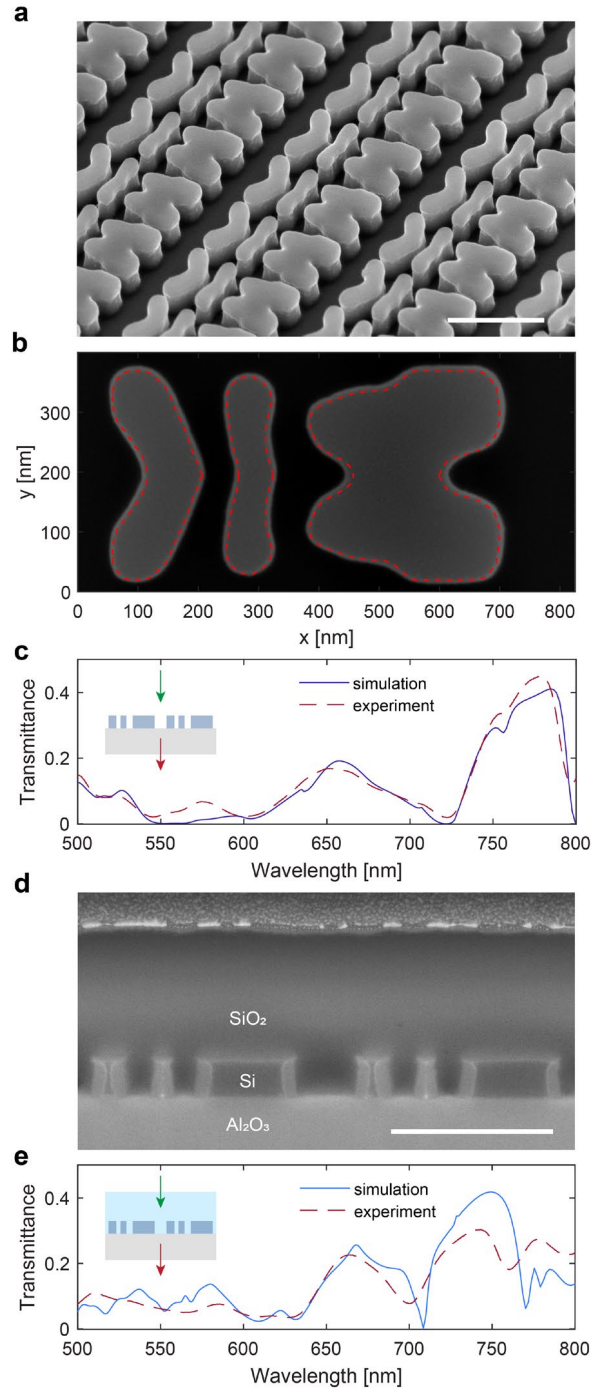

Fig. 4 | **Fabrication and optical characterization of the optical analog integral equation solver. a** Tilted scanning electron microscopy (SEM) image of the patterned Si metagrating. **b** Zoomed-in top-view SEM image of a single unit cell compared with the desired optimized contour (red dashed line). **c** Experimental (dashed red) and simulated (dark blue) transmittance spectra of the metagrating after the Si etch step (*i.e.* without silica spacer and top mirror). The sample was illuminated under normal incidence and the transmitted $0^{th}$-order diffraction intensity is collected (inset) as a function of incident wavelength. **d** SEM image of a focused-ion-beam-milled cross-section of the same metagrating embedded in a $SiO_2$ spacer. **e** Experimental (dashed red) and simulated (blue) transmittance spectra of the metagrating after the $SiO_2$ planarization step. The scale bar is 500 nm for panels a-d. The input polarization is TE for panels c-e.

First, we analyze the fabricated Si metasurface without the SiO$_2$ spacer and the semi-transparent mirror. As shown in Figure 4a, the fabricated structures after the Si reactive ion etching step are uniform and smooth over a large area. Next, it is important to compare the optimized unit cell to the experimental one. Figure 4b shows that the etched unit cell follows very closely the optimized contour (red dashed lines in Fig. 2b and Fig. 4b). To corroborate this feature, the transmittance of normal incident light to the 0$^{th}$ diffraction order was measured and compared to its simulated counterpart. In the simulation, the optimized structure described above was used. The transmittance spectrum was acquired over a broad wavelength range ($\lambda_0$=500-800 nm) to obtain maximum sensitivity in the comparison between experiment and simulation. Figure 4c demonstrates strong agreement between simulated and measured optical spectra, further confirming the suitability of the combined EBL+RIE process to fabricate precisely tailored metagratings for analog optical computing in the visible spectral range.

Next, the cross-section in Fig. 4d shows how the SiO$_2$ spacer conformally embeds the metagrating with no detectable air inclusions, creating a smooth planar top surface. The final thickness of the layer with the embedded metagrating amounts to 638 nm. Again, the transmittance is measured at this step, once more experimentally reproducing the key features present in the simulated ideal spectrum (see Fig. 4e). The small discrepancies between experiment and simulation in Fig. 4c-e can be attributed to minor fabrication imperfections, such as a slight difference in the SiO$_2$ refractive index between experiment and simulation, unintended resist over- or underexposure, and non-perfectly straight Si etching. Finally, the Au film evaporation concludes the fabrication, providing the metastructure with a semi-transparent mirror, and hence the required feedback system.

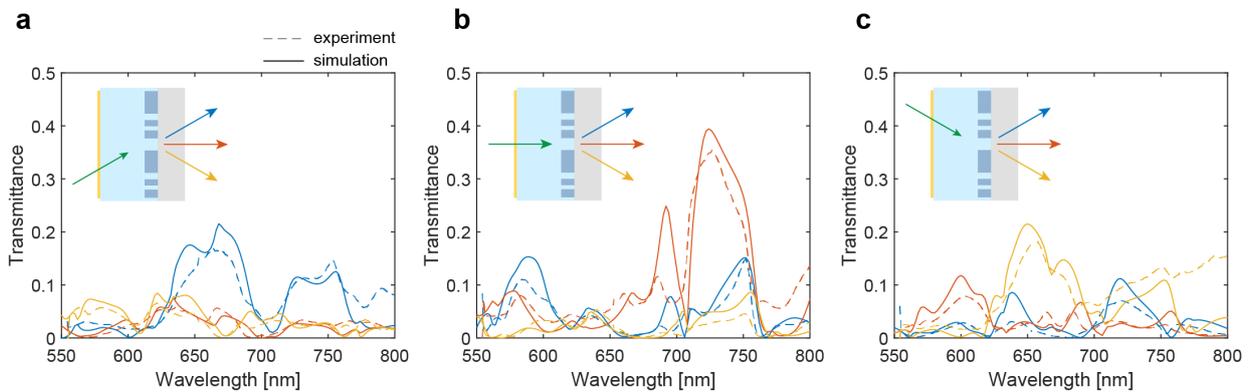

Fig. 5 | **Optical characterization of $S_{2T}$**. **a-c** Experimental (dashed lines) and simulated (solid lines) transmittance spectra of the completed metastructure. Insets: schematic visualization of the metastructure indicating the exciting input port (green arrows) representing orthogonal unit vectors, and the three output ports (yellow, orange, and blue matching the corresponding spectra). The input polarization is TE for all panels.

Figure 5 shows the measured transmittance spectra of the fully fabricated metastructure relating to each S-parameter belonging to $S_{2T}$ or, equivalently, the fraction of transmitted light going into each diffraction channel when the metastructure is illuminated through each input channel above the mirror. Specifically, each sub-panel shows the amplitudes squared of the elements belonging to each column of $S_{2T}$. Note that the input and output angles are changing with the wavelength of illumination according to the grating equation (see Supplementary materials for more information on the optical setup used). Each sub-panel in Fig. 5 also shows the simulated spectra of the designed ideal metastructure in Fig. 3a that gives the solutions shown in Fig. 3b.

The agreement over a broad wavelength range between simulation and experiment is clear: for each matrix element, the spectral features present in the simulation are reproduced experimentally. Small discrepancies between experiment and simulation are attributed to minor fabrication imperfections, as described above. Finally, taking advantage of the broad wavelength range of the data and minor perturbations to the structure in the simulation, it is possible to retrieve an estimate for the experimental solution provided by the fabricated metastructure, including its uncertainty (see Supplementary materials). Figure 6 compares the latter experimental solution to the ideal solution of Eq. (2) for the canonical inputs $(1,0,0)^T, (0,1,0)^T, (0,0,1)^T$. Although the accuracy of the solution is reduced compared to that shown in Fig. 3b and the wavelength of operation is blue-shifted by 7nm, the good agreement and similar trend with the ideal solution demonstrates the all-optical integral equation solving concept experimentally.

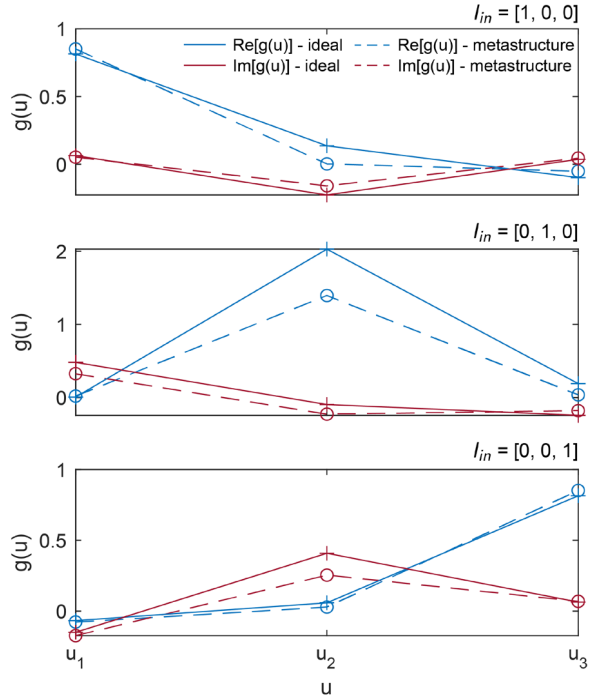

fig. 6 | **Estimated experimental solution**. Estimated experimental analog solution (real and imaginary parts) of the integral equation (dashed line – see Supplementary material) compared with the ideal theoretical solution. The wavelength of operation is $\lambda_0$ = 699 nm.

In conclusion, we have presented a Si-based optical metastructure that solves Fredholm integral equations of the second kind in a fully analog fashion at optical frequencies. First, we stated the mathematical problem in terms of Neumann series successive approximation. Next, we discussed the analogy between the integral equation solving and the behavior of an optimized periodic metagrating coupled to a feedback system. At the foundations of this mapping lies the possibility of designing the S-matrix of a periodic structure by setting its periodicity (i.e., the number of input/output modes and hence the dimension of the S-matrix) and optimizing its unit cell (i.e., optimize the coupling of light into the defined diffraction modes in amplitude and phase). Consequently, we showed how the designed metastructure effectively solves the problem of interest and compared the metasurface-based solution to the ideal solution. We showed that electron beam lithography and reactive ion etching provide the deep subwavelength spatial resolution required to create a hardware representation of a predefined Kernel, with relatively small deviations between experiment and simulations. We optically characterized the output for different input signals showing good agreement with the ideal simulated response.

Our results demonstrate the possibility of solving complex mathematical problems and a generic matrix inversion at speeds that are far beyond those of the typical digital computing methods. Our solution converges (within 3% difference from the infinite sum) in less than ten passes, corresponding to a processing time of about 20 fs, five orders of magnitude faster than the clock speed of a conventional processor. Operation in the visible spectral range in combination with deep-subwavelength fabrication resolution creates metastructures that are sub-micron thick. This represents a very high degree of circuit integration given the complexity of the mathematical operation performed in this small volume.

Further extensions of this work may explore non-symmetric kernels in a transmissive setup. Also, a similar scheme could be used to estimate eigenvalues of an integral operator (a matrix in the discretized form) by exploiting, for example, spontaneous emission of active materials. In fact, this would correspond to the solution of the homogeneous equation corresponding to the Fredholm integral equation of the second kind under study.

Moreover, it is possible to scale up the dimensionality of the problem, increasing the number of input/output ports by using more diffraction orders or by encoding information in the polarization state of light. The main challenge in either of these lies in one's ability to accurately fabricate unit cells of higher resolution and smaller feature size required to independently control more degrees of freedom. Of course, a larger number of diffraction orders would imply a larger periodicity and unit cell, partially easing this burden. One key advantage of our scheme is the possibility of integrating many designs within a unique feedback system, thus enabling parallelization whenever this is compatible with the problem under study.

Future applications of these concepts could include nonlinear materials within the feedback system (e.g., replacing the $SiO_2$ spacer layer) to explore nonlinear mathematical problems. Additionally, nonlinearity could also be applied after processing the information via linear operations. Hence, a dedicated external nonlinear device could be designed to process the outputs of our metastructure.

Finally, switchable metagratings (e.g. using phase change materials or mechanical modulation) could be envisioned to dynamically tune the encoded mathematical operation, paving the way for all-optical reconfigurable computing circuitry solving problems of further enhanced complexity.

**Acknowledgments**

This work is part of the research program of the Dutch Research Council (NWO) and is supported by the AFOSR MURI with Grant No. FA9550-17-1-0002. V.N.'s effort is supported by the National Science Foundation (NSF) MRSEC program under award No. DMR-1720530.

**Author contributions**

AC designed and fabricated the samples, performed numerical simulations and performed optical measurements. AC, BE and VN performed theoretical analyses. AA, NE and AP supervised the project. All authors contributed to the analysis and writing of the paper.

**ORCID**

Albert Polman 0000-0002-0685-3886

Andrea Alu   0000-0002-4297-5274

Andrea Cordaro 0000-0003-3000-7943

Brian Edwards 0000-0001-9354-125X

Nader Engheta 0000-0003-3219-9520